\documentclass[a4paper]{article}
\usepackage{graphicx}
\usepackage{onecolpceurws2018}
\usepackage{cite}
\usepackage{amsmath,amssymb,amsfonts}
\usepackage{algorithmic}
\usepackage{graphicx}
\usepackage{textcomp}
\usepackage{graphicx}
\usepackage{blindtext}
\usepackage{graphicx}
\usepackage{graphicx}
\usepackage{subcaption}

\graphicspath{{/home/ankit/Downloads/}}

\title{SFA-GTM: Seismic Facies Analysis Based on Generative Topographic Map and RBF}

\author{
Jatin Bedi \\ Dept. of CSE\\
                Indian Institute of Technology\\Roorkee, India 
\and
Durga Toshniwal \\ Dept. of CSE\\
                Indian Institute of Technology\\Roorkee, India 
}

\institution{}

\begin{document}
\maketitle

\begin{abstract}
Seismic facies identification plays a significant role in reservoir characterization. It helps in identifying the various lithological and stratigraphical changes in reservoir properties. With the increase in the size of seismic data or number of attributes to be analyzed, the manual process for facies identification becomes complicated and time-consuming. Even though seismic attributes add multiple dimensions to the data, their role in reservoir characterization is very crucial. There exist different linear transformation methods that use seismic attributes for identification, characterization, and visualization of seismic facies. These linear transformation methods have been widely used for facies characterization. However, there are some limitations associated with these methods such as deciding the width parameters, number of clusters, convergence rate etc. Therefore, the present research work uses non-linear transformation approach that overcomes some of the major limitations of linear approaches. The proposed Seismic facies analysis approach based on Generative Topographic Map \& Radial Basis Function(SFA-GTM) works by calculating the set of four Gray Level Co-occurrence Matrix(GLCM) texture based attributes viz. energy, homogeneity, contrast and dissimilarity. The Generative Topographic Map(GTM) is used for unsupervised classification of seismic facies based on the set of calculated texture attributes. Further, the present work uses Radial Basis Function(RBF) for interpolating the missing values in the data. 
\end{abstract}
\vskip 32pt

\section{Introduction}
Time series data\cite{1} is defined as a sequence of events measured over the repeated intervals of time. Time series mining has a number of applications in the field of stock market analysis\cite{2}, sales forecasting\cite{3}, weather forecasting\cite{3}, reservoir characterization\cite{4}, quality control\cite{4} and in many other areas. Seismic data is presented as a time series data with three spatial dimensions. Raw seismic data contains a set of traces. Seismic data\cite{bookseismic} helps to determine the structure of reservoir, but it plays a little role in determining the various properties in the reservoir bodies.
The act of building a reservoir model that incorporates all the characteristics of the reservoir to store hydrocarbons and also to produce them is called as reservoir characterization\cite{bookseismic,6}. Reservoir characterization is one of the biggest challenges that geophysicists face today. Well boring is one of the solutions to make the reservoir characterization better. However, it is an expensive and hazardous task for both humans and environment. It requires much time and makes the earth core unstable. Another way to make reservoir characterization is to do the efficient seismic facies classification\cite{6}. The basic aim of the seismic facies classification is to identify the rock properties and lithological changes in the facies\cite{bookseismic}.

Seismic face\cite{bookseismic} is defined as a region that exhibits some properties which distinguish it from the other areas. These variations of the properties in seismic facies would help in efficiently identifying the presence of various hydrocarbons in the seismic volume. Although there exist manual processes of analysing the seismic data in order to find out the region of interest. However, these manual methods are very time consuming and requires much expert knowledge to make accurate characterization\cite{bookseismic}. The problem becomes more difficult as the size of data increases or the number of attributes to be analyzed increases. Therefore, various techniques were introduced to automate the process of seismic facies classification. However, most of the automated methods were based on finding the prescribed pattern that makes them unreliable with noisy data. Apart from the conventional well logs, seismic attributes are being widely used in both exploration \& reservoir characterization and routinely been integrated into the seismic interpretation process. 

This study focus on texture attributes-based classification\cite{7} of seismic facies. It works by extracting a set of texture attributes from seismic data and uses those attribute for facies classification. The set of attributes are chosen on the basis of their variability around the seismic area of study. The texture attributes that shows most variations are considered for the purpose of classification. The approach can be broadly partitioned into the five steps:\\
(1) Collecting and Pre-processing the data.\\
(2) Extracting a set of features from time series data.\\
(3) Interpolating the missing values of attributes.\\
(4) Using non-linear approach for unsupervised facies classification.\\
(5) Visualizing the results. \\

\section{Literature Review}
Brian P. West\cite{8} proposed an approach for the seismic facies classification.  It begins by constructing the set of polygons on the cross-section area selected from the seismic volume. Gray level co-occurrence matrix had been calculated for each of the selected polygons. The set of different texture attributes i.e. homogeneity, energy, entropy generated by using GLCM matrix was then used as an input for the artificial neural network to identify the seismic facies. 

H.Sabeti\cite{sabeti} proposed an approach for seismic facies interpretation using K-means clustering algorithm. The method worked by calculating a synthetic seismic cube and eight different seismic attributes were calculated from the calculated seismic cube using paradigm software. The application of the approach to real seismic data demonstrated the detection of natural changes in the model. However, the biggest problem with K-means clustering algorithm is of determining the optimal number of clusters.
So, Atish Roy\cite{atish} introduced an algorithm that used the Self-Organizing Maps(SOM) to classify the facies of the Mississippian Tripolitic Chert reservoir. SOM\cite{som} is a method for clustering the data using prototype vectors. The approach begins by using the SOM for clustering. Initially, the method chooses a large value to define numbers of clusters. In the subsequent iterations, data vectors were merged into a small number of clusters. Both the structural and texture attributes were used for the facies classification. Further, supervision was introduced by using the three average vectors obtained as a result of the unsupervised classification. Different average data vectors had different attributes as components. The classification was performed by comparing each of the samples with the generated average data vectors. \smallbreak

J.D.Pigott\cite{john} used first order seismic attributes for identification of seismic facies. There are numerous attributes available for interpretation of seismic facies. John D.\cite{john} defined the eight first order seismic attributes that play a major role in geological interpretations. Attributes identified were: Amplitude, Instantaneous Frequency, Variance, Chaos, Envelope, Acoustic Impedance and Cosine of phase. These eight attributes were then used for basin exploration of East China Sea.
Renjun Wen\cite{renjun} introduced a new approach for 3-D modelling of heterogeneity present in the channelized reservoirs. The method worked by calculating the acoustic impedance cube by using the deterministic seismic inversion. The set of six seismic attributes calculated from the impedance cube were then used for the purpose of seismic facies classification. The approach implements the two methods for seismic facies classification. First, Trace based and second voxel based classification. Trace based classification method worked by assigning facies code to each trace within a cube whereas in voxel based classification each small voxel is considered for classification. The classification algorithm used the six attributes calculated previously. Resulting cube obtained by using a different combination of attributes were then compared with the Ground truth. Based on the comparison, only the significant attributes were considered for classification. However, the approach did not perform well with the noisy data.\smallbreak

So, Hao-Kun Du\cite{hao} introduced an approach for seismic facies analysis using the Self-Organizing Maps (SOM) and Empirical Mode Decomposition(EMD). EMD is a method for de-noising the data. It works by calculating the IMFs (Intrinsic Mode Functions). IMFs that shows good correlation with the data is then used to represent the data. SOM works on the principle of unsupervised learning to present the data into lower-dimensions. SOM\cite{som} works in two phases which are
(1) Learning phase
(2) Mapping phase.
In the learning phase, SOM learns by using the input data and during mapping phase, each of the input vectors is mapped to one of the nodes in the SOM grid. 
The application of this approach to real seismic dataset shows that the facies generated were better than that of SOM without Empirical Mode Decomposition(EMD).\smallbreak

Tao Zhao in 2015\cite{last} performed a comparison of various supervised and unsupervised seismic facies classification techniques including K-means, SOM, GTM, Gaussian Mixture Model(GMM), Support Vector Machine(SVM) \& Artificial Neural Network(ANN). These six classification algorithms were applied to 3-D seismic data volume acquired over the Canterbury Basin, New Zealand. The classification results obtained by the application of various algorithms to the seismic data of study area shows that supervised methods provide accurate estimates of the seismic facies. However, they fail to identify some of the important features highlighted by the unsupervised methods.\smallbreak 

As pointed out above, Seismic facies classification using K-means algorithm\cite{sabeti} identifies the literal changes in the facies, but fails to determine the optimal numbers of clusters. Further, the facies classification using Self-Organizing Maps(SOM)\cite{atish} introduced by Hao-Kun du also suffers from some limitations including deciding optimal parameters, convergence, etc. All the approaches discussed above performs a linear transformation from data space to latent space. Therefore in the present work, we introduce a classification(SFA-GTM) approach that not only performs the non-linear transformation from data space to latent space but also removes the major limitations of the existing approaches. The rest of the paper is organized as follows: an introduction to Generative Topographic Map(GTM) and data set description are given in section 3 \& 4 respectively. The methodology of the proposed approach for seismic facies classification is specified in section 5. Section 6 explains the results of classification \& conclusion is stated in section 7.  

\section{Introduction to GTM}
The principal component analysis\cite{pca} is defined as a method for projection of a point from M-dimensional space into a hyperplane of L-dimensions where L $\leq$ M. Similarly the factor analysis\cite{fa} is also used for representing a large number of correlated variables into a small number of factors that better represents the data. The difference between the two is that factor analysis deals with the co-variance whereas the PCA focuses on the variance in data\cite{fpca}. Factor analysis is based on the EM\cite{em} (Expectation-Maximization) algorithm and it works by modeling M-dimensional variable as a function of latent variables(L-Dimensional) plus the noise. 
\begin{figure}[htbp]
\centering
\includegraphics[width=10cm,height=7cm]{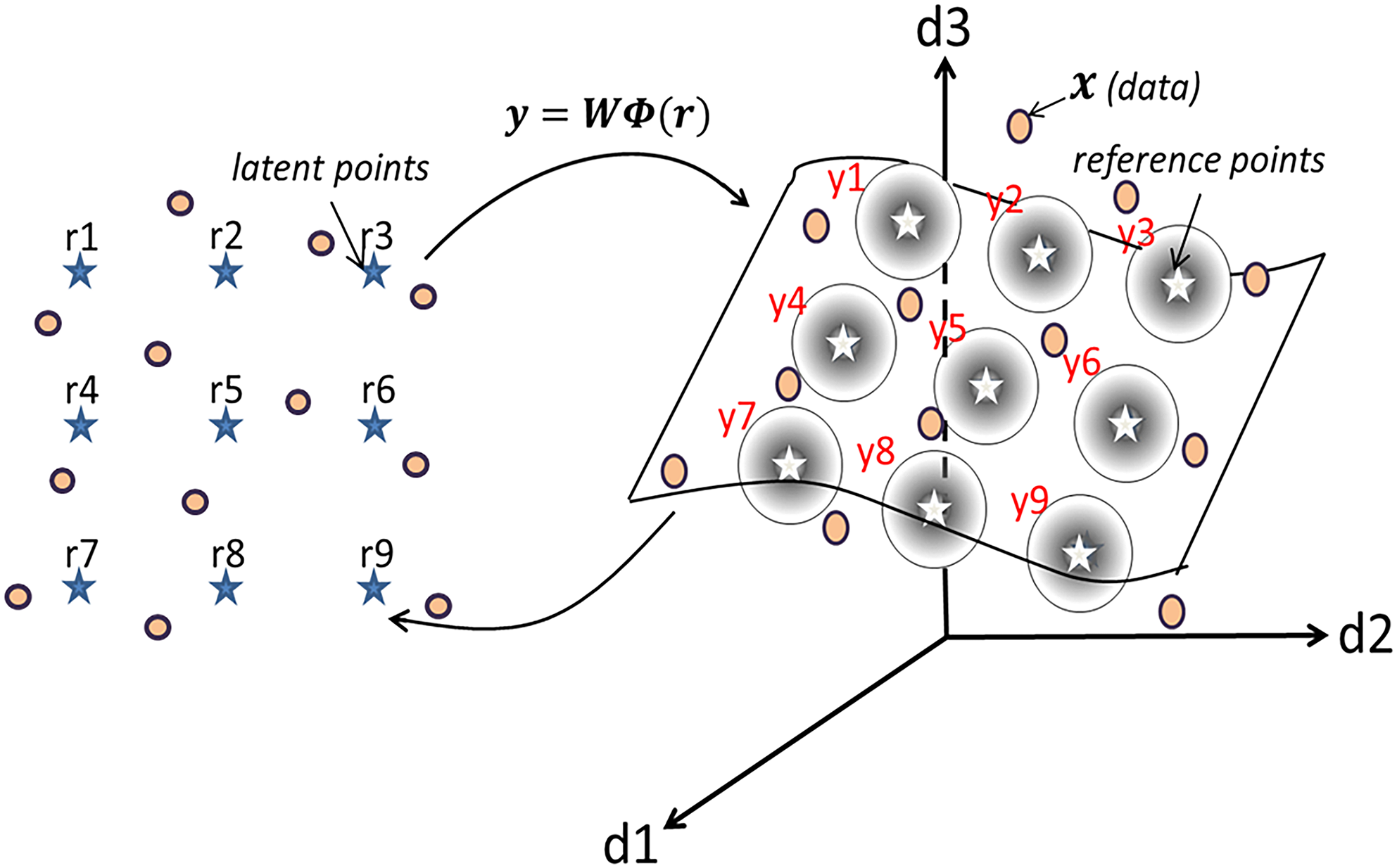}
\label{An Example of Non-Linear Transformation from Data Space(3-Dimensions) to Latent Space(2-Dimensions)}
\caption{An Example of Non-Linear Transformation from Data Space (3-Dimensions) to Latent Space (2-Dimensions) by embedding a manifold in data space\cite{figurecitation}, where {$r_{1},r_{2}....r_{n}$ denotes the points in latent space, $y=W.\phi(r)$ represents the mapping function.}}
\end{figure}
Further, SOM(Self-Organizing Maps)\cite{som} is also a method for low-dimensional representation of input vector space. There are three essential processes in the formation of SOM i.e. Competition, Co-operation, and Synaptic Adaptation. In Competition phase, each neuron computes discriminant function for input vector. The neuron with the largest discriminant wins the competition. In Co-operation phase wining neuron excites the adjacent neurons depending upon their distance from winning neuron. Synaptic adaptation phase enables the adjacent neurons to increase their discriminant function in response to the event. \smallbreak

Although SOM has been widely used in various application such as image processing, visualization, facies classification \& unsupervised learning but there are various limitations associated with it. Firstly, there does not exist any framework by using which the SOM initial parameters can be chosen i.e. learning rate, neighbour function width etc. Second, there is no assurance/promise that the training algorithm will converge.\\

Like Self-Organizing Maps, Generative Topographic Map (GTM)\cite{gtm} represents the input data into a small number of latent variables. GTM removes some of the major drawbacks of SOM. It is a form of non-linear transformation from the input data space to the latent space. The conversion between the two spaces is done with the help of a function $ \phi(x; W)$, where $x$ denotes a point in the latent space and $W$ denotes the weight and bias values. The mapping function inserts a Non-Euclidean manifold onto the data space as shown with the help of an example with latent space dimension(L) equals to 2 \& data space dimension(D) equals to 3. Introducing a distribution over the latent space leads to the generation of a probability distribution over the data space.

In Generative topographic map, we consider the prior distribution over the latent space to be Gaussian\cite{gtm} (for given $x$ and $W$) given by equation (\ref{eq-1}) :
\begin{equation}
 p(t|x,W,\beta)=(\frac{\beta}{2\pi})^{D/2} \exp\{(-\frac{\beta}{2}\Arrowvert\phi(x;W)-t\Arrowvert^{2}\} \label{eq-1}
\end{equation}
 
where $\beta$ is noise variance and $t$ is a point in data space.\\
The integration over latent space distribution will result into the induction of posterior distribution\cite{gtm} over the data space and is given by equation (\ref{eq-2}).
\smallbreak
\begin{equation}
p(t|W,\beta)= \int p(t|x,W,\beta)p(x) dx \label{eq-2} \\
\end{equation}

After determining the prior distribution and the mapping function the initial value of $\beta$ and $W$ are determined by using principal component analysis\cite{pca}. However the integration over $x$ in Equation 2 is analytically intractable. The model can take the different form depending on the function $\phi(x;W)$. In order to make the model similar in powers/spirit to SOM we use a special form of delta functions centred on the nodes of grid in latent space\cite{gtm1} given by equation (\ref{eq-3}):
 \begin{equation}
 p(x)= 1/K \sum\limits_{i=1}^{K} \delta(x-x_{i}) \label{eq-3} \\
 \end{equation}

By using this form of $p(x)$ the integration given in equation (2) can be performed analytically. Each point mapped by using mapping function forms the center of a Gaussian density function. So the distribution function in data space becomes 
\begin{equation}
p(t|W,\beta)=1/K \sum\limits_{i=1}^{K} p(t|x_{i},W,\beta) \label{eq-4}\\
\end{equation}
In general the objective function in GTM algorithm takes the form of log likelihood which is given by equation (\ref{eq-5}):
\begin{equation}
\mathcal{L}= \sum\limits_{n=1}^{N} Ln\{1/K \sum\limits_{i=1}^{K} p(t_{n}|x_{i},W,\beta)\} \label{eq-5}
\end{equation}

\subsection{Expectation-Maximization Algorithm}

After deciding the mapping function \& initial value of $W$ and $\beta$ we use the expectation Maximization\cite{em} algorithm for non-linear optimization. Consider at some point in algorithm we have weight matrix $W_{old}$ and variance $\beta_{old}$. The E-step in EM algorithm proceeds by computing the responsibility for each combination of k and n, where k is the Gaussian component and n is the data point\cite{gtm1}.\\
The responsibility value $r_{kn}$ denotes the posterior probability that $n^{th}$ data point was generated by $k^{th}$ component and is given by equation (\ref{eq-6})\\
\begin{equation}
r_{kn}=p(x_{k}|t_{n},W_{old},\beta_{old})=\frac{p(t_{n}|x_{k},W_{old},\beta_{old})}{\sum\limits_{i=1}^{K}p(t_{n}|x_{k},W_{old},\beta_{old})} \label{eq-6}
\end{equation}
Usually, the mapping function used in topographic mapping is linear regression model where $\mathcal{Y}$ is a linear combination of basis functions given by:
\begin{equation}
\mathcal{Y}=\sum \delta* W \label{eq-7}
\end{equation}
The M-step is used to calculate the new updated values of weights $W$ \& $\beta$ by derivating log-likelihood ($\mathcal{L}$) w.r.t $W$ and setting it to zero. GTM algorithm alternate between these two E and M step until objective function is converged\cite{gtm1}.
\begin{table}[htbp]
\caption{Survey Parameters}
\label{tab}      
\centering
%% Some packages, such as MDW tools, offer better commands for making tables
%% than the plain LaTeX2e tabular which is used here.
\renewcommand{\arraystretch}{1.5}
\begin{tabular}[5cm]{|p{3cm}|p{4cm}|}
\hline
$\textbf{Parameter}$ & $\textbf{Range of Parameter}$\\
\hline
$\textbf{Inline Range}$ & $[100-750] $\\
\hline
$\textbf{Crossline Range}$ & $[300-1250]$\\
\hline
$\textbf{Z Range}$ & $[0-1848]$\\
\hline
$\textbf{Size(km)}$ & $[24*16]$\\
\hline
\end{tabular}
\end{table}
\section{Dataset Description}
The reservoir characterization needs two types of data: hard and soft data. This study uses the data of seismic survey conducted in the block of Dutch sector of North Sea\cite{gt}. It covers an area of approximately 24*16 $km^{2}$. The data set is collected from the dGBEarthSciences. The seismic data volume comprises of 947 cross-lines and 646 in-lines. Both in-lines and cross-lines have a line spacing of 25 m and the sample rate is 1 ms. The complete description of survey parameters is given in Table \ref{tab}.
\section{Proposed Work}
Facies Classification is the process of assigning a label to each of face depending on its distinct properties. There are two types of learning\cite{ml} supervised \& unsupervised. Supervised algorithms\cite{ml} builds a learning model by using the training data and employs that learning model to determine the output label for test data. While for unsupervised algorithms\cite{ml} no training data are made available. The seismic facies identification is one of the main problems in the area of reservoir characterization. There exist various unsupervised approaches for seismic facies identification such as Self-Organizing Maps, k-means clustering etc. A number of recent studies were based on the facies classification using these approaches. However, there are some limitations associated with these approaches such as convergence problem, lack of a theoretical framework for choosing model parameters etc. Therefore in this paper, we introduce a non-linear approach(SFA-GTM) for unsupervised classification of seismic facies based on the set of attributes. The classification is called unsupervised because no well data is used in the proposed approach. The proposed approach solves some of the major limitations of already existing approaches. Further, the new approach also introduces an interpolation method that is used in conjunction with GTM for filling the missing values of attributes in the seismic data. \figurename{2} demonstrates the flowchart of the proposed SFA-GTM approach for facies classification.
\subsection{Input Seismic Data}
The very first step in any data mining task is collecting the data. Seismic data is collected by means of seismic surveys\cite{bookseismic}.
In seismic surveys, the waves generated by a set of seismic sources passes into the earth and the rays that reflect back to the surface are recorded by seismic sensors. The time taken by the different rays to reach the sensors provides valuable information about the rock types and any possibility of gas and fluid in rock formations\cite{bookseismic}.
\begin{figure}[htbp]
\centering
\includegraphics[width=14cm,height=2.9cm]{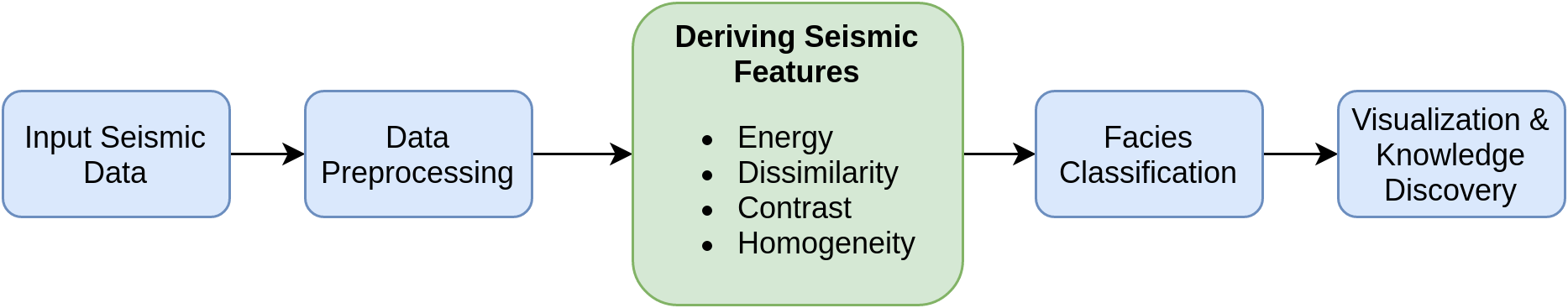}
\label{An Example of Non-Linear Transformation from Data Space(3-Dimensions) to Latent Space(2-Dimensions)}
\caption{Proposed Methodology}
\end{figure} 
\subsection{Data Pre-processing}
Data pre-processing plays a vital role in nearly all accurate analysis. The raw data collected from sensors is in "unprocessed" form. It suffers from many limitations including noise, dead traces, etc. The classification accuracy highly depends on the quality of data used. Therefore, there is a rising need of pre-processing the collected data to make the accurate facies analysis. This study uses the Dip-stirred filtering\cite{filter} to remove the noise present in 3-D seismic data. Even though the data pre-processing makes the data suitable for the analysis, but it is an expensive and time-consuming task.
\begin{figure*}[!t]
\centering
\begin{subfigure}{0.5\textwidth}
\centering
\includegraphics[width=8cm,height=4.8cm]{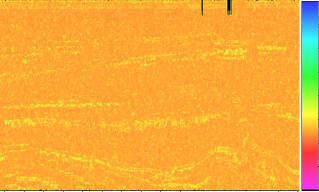}
\caption{GLCM Energy}
\label{fig:left}
\end{subfigure}
\begin{subfigure}{0.49\textwidth}
\centering
\includegraphics[width=8cm,height=4.8cm]{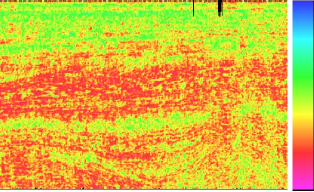}
\caption{GLCM Contrast}
\label{fig:right}
\end{subfigure}
\begin{subfigure}{0.5\textwidth}
\centering
\includegraphics[width=8cm,height=4.8cm]{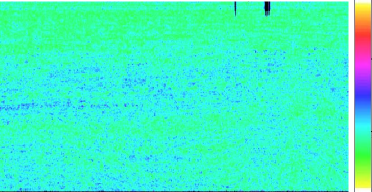}
\caption{GLCM Homogeneity}
\label{fig:left}
\end{subfigure}
\begin{subfigure}{0.49\textwidth}
\centering
\includegraphics[width=8cm,height=4.8cm]{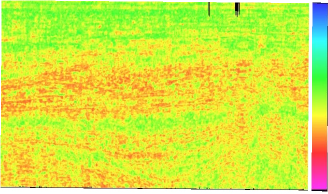}
\caption{GLCM Dissimilarity}
\label{fig:right}
\end{subfigure}
\caption{GLCM Texture Attributes}
\label{fig:combined}
\end{figure*}
\subsection{Calculation of Set of Seismic Attributes}
Seismic attributes play a major role in the interpretation of seismic data\cite{6}. Attributes make the characterization of rock properties and lithological changes easier and better. This study performs the GTM based classification of seismic facies based on the set of GLCM-texture attributes\cite{texture}.

The texture attributes are calculated by means of Gray Level Co-occurrence Matrix(GLCM)\cite{texture}. It assumes the set of traces in the seismic data as an image. Gray level matrix depends upon the organization of pixel values. The frequency with which two-pixel values occurs together is output from the GLCM function\cite{texture}. The function works by generating a M*M matrix where m denotes the number of the gray level scale used. For this paper, we consider the matrix to be of 64*64. There exist some different GLCM seismic attributes named mean, variance, contrast, entropy, dissimilarity, homogeneity, energy etc. for facies analysis. However, this paper chooses the four textural attributes depending upon the degree of correlation between them. The approach proceeds by calculating the four textural attributes named Energy, Homogeneity, Dissimilarity and Contrast. Fig.\ref{fig:combined} shows the four texture attributes calculated for the seismic data. 

GLCM contrast is basically used to determine any local changes in the data and is given by equation (\ref{eq-8})\cite{attributes}.
\begin{eqnarray}
\mathcal{C}= \sum\limits_{i,j=0}^{N} P_{i,j}(i-j)^2 \label{eq-8}
\end{eqnarray}

Where P is the GLCM probability matrix and N is the size of the matrix.

\smallskip
GLCM Energy\cite{formula} is directly related to the cooccurrence matrix and express the continuity of rock properties. It determines the presence of homogeneous or rough texture and is given by equation (\ref{eq-9})

\begin{equation}
\mathcal{E}= \sqrt{\sum\limits_{i,j=0}^{N-1} P_{i,j}^2} \label{eq-9}
\end{equation}
\smallskip
The low value of energy corresponds to rough texture and high to the homogeneous texture\cite{attributes}.

\smallskip
GLCM homogeneity\cite{formula} is used to determine the smoothness in the texture i.e degree of neighbourhood similarity in the data. It identifies the region with static mean and variance and is given by equation (\ref{eq-10})
\begin{equation}
\mathcal{H}=\sum\limits_{i,j=0}^{N-1} \frac{P_{i,j}}{1+(i-j)^2} \label{eq-10}
\end{equation}

GLCM dissimilarity\cite{formula} does the comparative analysis in the nearby regions and is given by (\ref{eq-11})

\begin{equation}
\mathcal{D}=\sum\limits_{i,j=0}^{N-1} P_{i,j} \Arrowvert i-j \Arrowvert \label{eq-11}
\end{equation}
\begin{figure*}[!t]
\centering
\begin{subfigure}{1\textwidth}
\centering
\includegraphics[width=12.5cm,height=4.8cm]{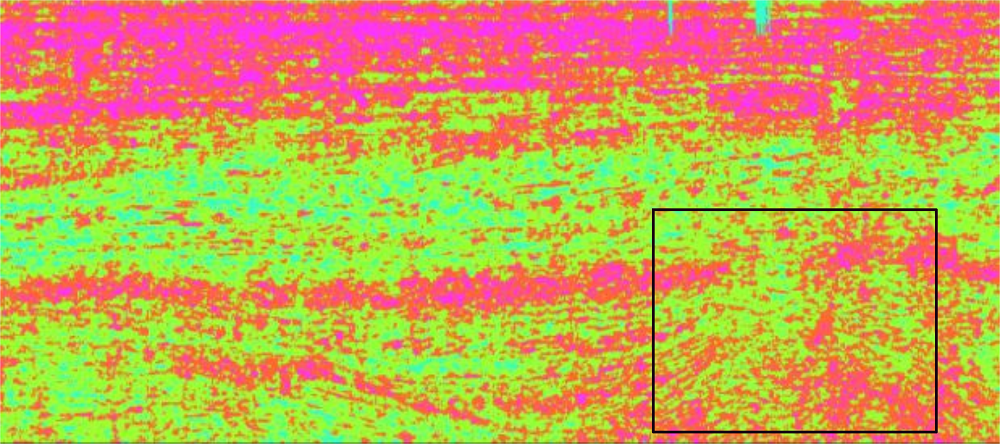}
\caption{Facies Classification using Linear Transformation Approach}
\label{fig:left1}
\end{subfigure}
\begin{subfigure}{1\textwidth}
\centering
\includegraphics[width=12.5cm,height=4.8cm]{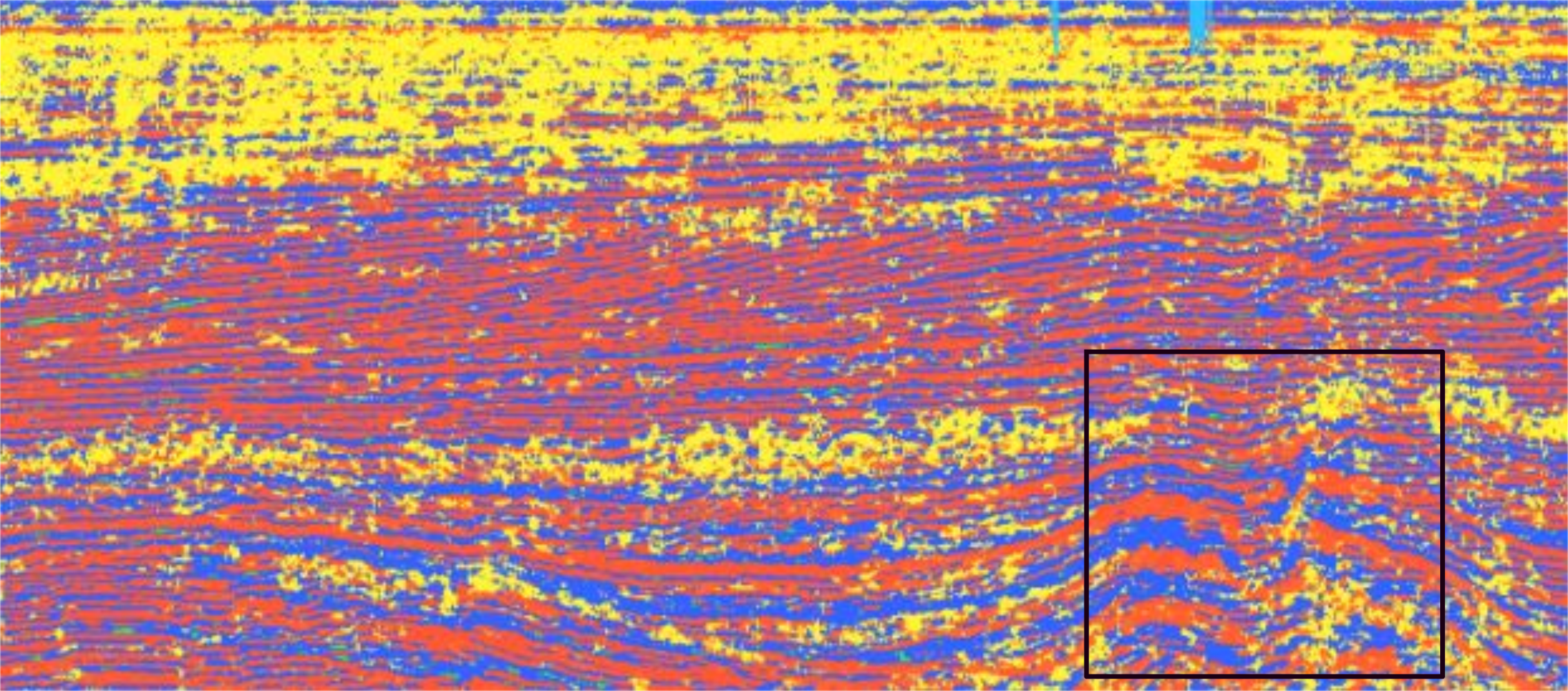}
\caption{Facies Classification using Non-Linear Transformation(GTM) Approach}
\label{fig:right1}
\end{subfigure}
\caption{Visualizing Seismic Facies Classification }
\label{fig:combined1}
\end{figure*}
\subsection{Missing Values Handling}
The Pre-processing task recovers the major problems present in the data including removal of noise, dead traces etc. However, there can be the case where the degree of distortion in the data is too high to be recovered by means of the pre-processing task. The level of distortion in the data depends on the various factors including recorder settings, type of source devices used in the survey, the area of seismic survey etc. So there might be the chances that calculation of attributes over that part of the dataset would make the overall process inaccurate. The most common problem in nearly all fields is that we have values of variables available at certain locations and we want to find a function that uses the existing data to determine values at locations different from the measured locations. There exists some machine learning algorithms\cite{ml} for handling missing values in the data. In this study, we have used Radial Basis Function(RBF)\cite{rbf} for interpolating the missing values of attributes in the data. 
RBF has a number of applications in different fields including data mining\cite{rbf}, machine learning\cite{rbf1} and statistics\cite{rbf1}. The overall goal of using RBF for missing values prediction is to improve the accuracy of the facies classification process.  \\
Given a set S of input vector of D-dimensions and corresponding set $\mathcal{O}$ of output.
\begin{equation}
S=\{s_{i}^{m} : i=1,2,3,....D\}
\end{equation} 
The aim of RBF interpolation\cite{rbf1} is to find a function $\mathcal{F}$ such that
$\mathcal{F}$(S)=$\mathcal{O}$. The present work uses the RBF for filling the missing attribute values in the data. The degree of relationship between near well features and seismic attributes forms the basis for the RBF. The method works by dividing the complete labelled dataset into the train and test part. Train part is used to build a learning model. The learning model is built separately for each individual attribute. After proper training of learning models, we use them to predict the attribute values for test part of the dataset. The difference between actual and predicted value of attributes gives error value. The root mean square error values for the RBF models built in this study are given in Table \ref{tab2}. The Error Values are very low which shows that the models are developed well. These resulting RBF models are then used for predicting the attributes values for the missing part of the data.
\begin{table}[htbp]
\caption{RBF Results}
\label{tab2}      
\centering
%% Some packages, such as MDW tools, offer better commands for making tables
%% than the plain LaTeX2e tabular which is used here.
\renewcommand{\arraystretch}{1.5}
\begin{tabular}[5cm]{|p{4cm}|p{3cm}|p{3cm}|}
\hline
$\textbf{Name of Attribute}$ & $\textbf{Training Error}$ & $\textbf{Testing Error}$\\
\hline
$\textbf{Energy}$ & $0.178 $ & $0.213 $\\
\hline
$\textbf{Homogeneity}$ & $0.182 $ & $0.201 $\\
\hline
$\textbf{Contrast}$ & $0.203 $ & $0.238$\\
\hline
$\textbf{Dissimilarity}$ & $0.194 $ & $0.197 $\\
\hline
\end{tabular}
\end{table} 

\subsection{Facies Classification Using Generative Topographic Map }
  
The next step after calculating the four textural attribute energy, homogeneity, contrast, dissimilarity and missing value handling is to initialize the non-linear mapping model. In the present work, the aim is to perform a non-linear transformation keeping D=4 dimensional data space to L=2 Dimensional latent space. The model begins by initializing the 2-D grid of latent point with 30*30 nodes. Then the algorithm generates the grid of basis function centers of 15*15 nodes and also selects the values of sigma where sigma denotes the width of the basis function.

The initial values of the parameters (Weights $\mathcal{W}$ and Bias $\mathcal{(B)}$) required for initializing the Mapping model is assigned by using the principal component analysis algorithm\cite{pca}. The non-linear model then proceeds by computing the prior probability over the latent space. As pointed out in section (III), defining a prior distribution over the latent space will induce the corresponding distribution over the data space. The algorithm goes on updating the values of Weights $\mathcal{W}$ and Bias $\mathcal{(B)}$ by executing the Expectation-Maximization(EM) algorithm till it fulfills stopping criteria, where stopping criteria would be decided by two factors first when the error gets constant and second when the iteration reaches the maximum number of units.

\section{Classification Results}
The non-linear generative topographic map performs a transformation of data from four-dimensional seismic attributes plane to two-dimensional seismic facies plane for visualizing the classification results. Fig. \ref{fig:combined1} shows the facies classification resulting two transformation approaches. The Fig. \ref{fig:left1} presents the classification results obtained from the linear transformation approach\cite{fpca} and Fig. \ref{fig:right1} demonstrates the facies classification resulting from the Non-linear GTM.
As per the classification results shown in Fig. \ref{fig:combined1}, four different kinds of facies are identified by the generative topographic mapping. From the Fig. \ref{fig:combined1} it is quite clear that the facies identified by using generative topographic map are more accurate than the facies identified by linear transformation approach. The comparison of regions of seismic facies highlighted with the rectangle in Fig. \ref{fig:left1} \& \ref{fig:right1} shows that the facies identified by non-linear approach are more clearly identifiable as compared to that of linear approach. Moreover, the comparison of different seismic regions identified by using GTM with the ground truth\cite{gt} shows that the approach performed an accurate facies classification.

\section{Conclusion} 
In this paper, we presented a nonlinear transformation approach(SFA-GTM) to identify different seismic facies. The proposed SFA-GTM approach serves the two major purposes. Firstly, it provides a method for interpolating the missing entries in the data. Radial Basis Function(RBF) is used for interpolating the missing values of attributes. The Root Mean Squared(RMS) error values obtained as result of interpolation shows that the method accurately interpolates the missing values of attributes in the data. Secondly, the proposed approach performs a nonlinear mapping between data space and latent space. There exist various linear approaches such as SOM, K-Means to identify seismic facies. However, these linear approaches suffer from some major limitations such as determining convergence rate, natural clusters, width parameters etc. In contrast to linear transformation techniques, GTM solves the biggest problem of convergence by determining the value of stability of variance. The proposed SFA-GTM approach identifies the seismic facies by using four textural attributes. The approach works well even in the absence of well log data and identifies the natural clusters present in the data. Furthermore, the classification result shows that the set of facies identified by the GTM is more precise than linear approaches. However, the major limitation of the proposed approach is that it has higher run-time complexity than linear approaches. This study is primarily focused on the unsupervised classification of seismic facies. The non-linear supervised classification of facies using seismic and well log data can be taken as future work.

\end{document}